\documentclass[aps,prl,twocolumn,showpacs,floatfix,superscriptaddress]{revtex4}

\usepackage{graphicx}
\newcommand {\be}{\begin{equation}}
\newcommand {\ee}{\end{equation}}
\newcommand{\deff}{{\mathcal D}_{\rm eff}}
\newcommand{\tlock}{\langle T_{lock}\rangle}

\begin{document}

\title{Experimental Study of Noise-induced Phase Synchronization 
     in Vertical-cavity Lasers}

\author{Sylvain Barbay}
\affiliation{Laboratoire de Photonique et de Nanostructures, 
CNRS-UPR 20, Route de Nozay 91460 Marcoussis, France}

\author{Giovanni Giacomelli}
\email{giacomelli@inoa.it}
\affiliation{Istituto Nazionale di Ottica Applicata, largo E. Fermi 6
50125 Firenze, Italy}
\affiliation{Istituto Nazionale di Fisica della Materia-UdR
Firenze, Via G. Sansone 1 50019 Sesto Fiorentino, Italy }

\author{Stefano Lepri}
\affiliation{Dipartimento di Energetica ``S. Stecco",
via S. Marta 3 50139 Firenze, Italy}
\affiliation{Istituto Nazionale di Fisica della Materia-UdR
Firenze, Via G. Sansone 1 50019 Sesto Fiorentino, Italy }

\author{Alessandro Zavatta} 
        
\affiliation{Dipartimento di Sistemi e Informatica,
via S. Marta 3 50139 Firenze, Italy}

\date{\today}

\begin{abstract}

We report the experimental evidence of noise-induced phase synchronization  in
a vertical cavity laser. The polarized laser emission is entrained with the
input periodic pump modulation when an optimal amount of white, gaussian noise
is applied. We characterize the phenomenon, evaluating the average frequency of
the output signal and the diffusion coefficient of the phase difference
variable. Their values are roughly independent on different waveforms of 
periodic input, provided that a simple condition for the amplitudes is
satisfied.  The experimental results are compared with numerical simulations of
a Langevin model.   

\end{abstract}


\maketitle


The phenomenon of Phase Synchronization (PS) has been the subject of extensive
investigation during the last years (see e.g. \cite{bookpiko} for a recent
review). In the purely deterministic case, two systems with instantaneous
phases $\Phi_1(t)$ and $\Phi_2(t)$ are locked if, for two integers $n,m$,
$n\Phi_1-m\Phi_2$ remains bounded for all times. In the presence of
unbounded fluctuations, this condition on the phase difference cannot hold for
all times. Sufficiently large noise  will eventually cause phase slipagge and
PS can thus only be effective.  For example, for weak periodic forcing of an
autonomous stochastic oscillator, the dynamics of the instantaneous phase
difference $\phi$ is ruled by the  Adler equation \cite{adle1946,stra1967} 
\begin{equation} 
\label{eq:stochadler}
\dot\phi =\Delta-\Delta_s\sin\phi + \xi 
\end{equation} 
where $\xi$ is a Gaussian white noise. Eq.~(\ref{eq:stochadler}) describes the
motion of an overdamped Brownian particle in the tilted potential $-\Delta
\phi-\Delta_s\cos\phi$. Therefore, if the frequency mismatch  
$\Delta$ is smaller than the synchronization bandwidth  
$\Delta_s$, locking can occur. However, from time to time
fluctuations suffice to kick $\phi$ out of the well yielding a phase slip.
Recent observations of noise--enhanced PS have been reported in more complex
situations, where chaos occours in absence of noise \cite{kurths02,bocca03}.
In this context, dynamical features may be significantly altered and 
a richer variety of phenomena arise \cite{report}.

For noise-driven bistable systems, PS is strictly related to the phenomenon of
Stochastic Resonance (SR), i.e., the enhancement in the response to a
weak signal superimposed to a stochastic input (see e.g.
\cite{rmp1998}). The very fact that noise can induce synchronized jumps between
the two stable output states has been recently interpreted as a noise-induced 
PS. A first observation of PS in an electronic circuit has been
reported in \cite{RUS}, analyzing the output mean frequency as a function of
added noise.  Numerical simulations of Langevin models \cite{neim1998,callen}
showed a similar behavior and allowed for a more refined description of PS in
terms of the effective phase diffusion coefficient \cite{stra1967}.   An
analytic theory, performed in the case of a periodic and aperiodic square wave
modulation, was presented by Freund et al. in \cite{freu2000}, where the
explicit expressions for the mean output frequency and the effective diffusion
coefficient were derived from  a Master Equation for the  stochastic variable
$\phi$. 

In this work, we present a detailed experimental investigation of noise-induced
PS regimes in a a bistable optical system. Our study is
motivated by the necessity to perform a throughful test of the statistical
indicators of PS in reliable, well controllable laboratory conditions. The effect
of both sinusoidal and square-wave input periodic modulations are considered; the
latter case allows to discuss a general framework for a comparison with the
analytic results of Ref.\cite{freu2000}.

Our experimental system is a pump-modulated Vertical Cavity Surface Emitting
Laser (VCSEL).  The VCSEL is a semiconductor laser, with a symmetrical cavity
allowing  for the emission on two linear, perpendicular polarizations selected
by the crystal axis directions (see e.g. \cite{laser}).  The emission symmetry,
usually broken by impurities or optical inhomogeneties leading to a
single-polarization emission, may be restored for particular choices of the
pump current. In such a case, the system is bistable and exhibits noise-driven,
random jumps between the two polarizations ruled by the Kramers' law
\cite{kram1940}. The experimental evidence of SR \cite{giac} and
of binary aperiodic SR \cite{barb2000} in a VCSEL has been recently reported. 

In our setup, we employ a VCSEL lasing at 850~nm, thermally  stabilized
(better than 1~mK) and with a carefully controlled pump current. The overall
stability allows for long--time measurements, even in presence of critical
behaviors. A  linear polarization direction in the laser emission is selected
using a polarizer and an half-wave plate. The laser intensity is monitored by
an avalanche detector  and the signal is recorded by a digital scope. An
optical isolator prevents from  optical feedback effects to occur. The signals
from a 10~MHz-bandwidth white-noise generator and a sinusoidal (SIN) or a
square--wave (SQR) oscillator are summed and coupled into the laser by means of
a bias--tee. The period of the input signals was chosen to be $T = 5\, \mu s$ 
(200~kHz) for both waveforms. The measurements are performed varying
the intensity $D$ of the applied noise as well as the 
amplitude $A$ of the input modulation.  For all the measurements reported
henceforth, the $A$ values are subthreshold, i.e. small enough to avoid output
transitions every modulation period (in absence of added noise).  
The amplitude $A_{SQR}$ is set to be equal to  the root mean
square value of $A_{SIN}$, i.e. $A_{SQR}=A_{SIN}/\sqrt{2}$. For every choice of
the parameters, a sequence of 16 million points is acquired, corresponding to
500 samples/period for a total of 32,000 periods. Such large data sets allow to
reduce statistical errors  to a level of accuracy comparable with current
numerical simulations of theoretical models.

\begin{figure}
\includegraphics[clip,width=8cm]{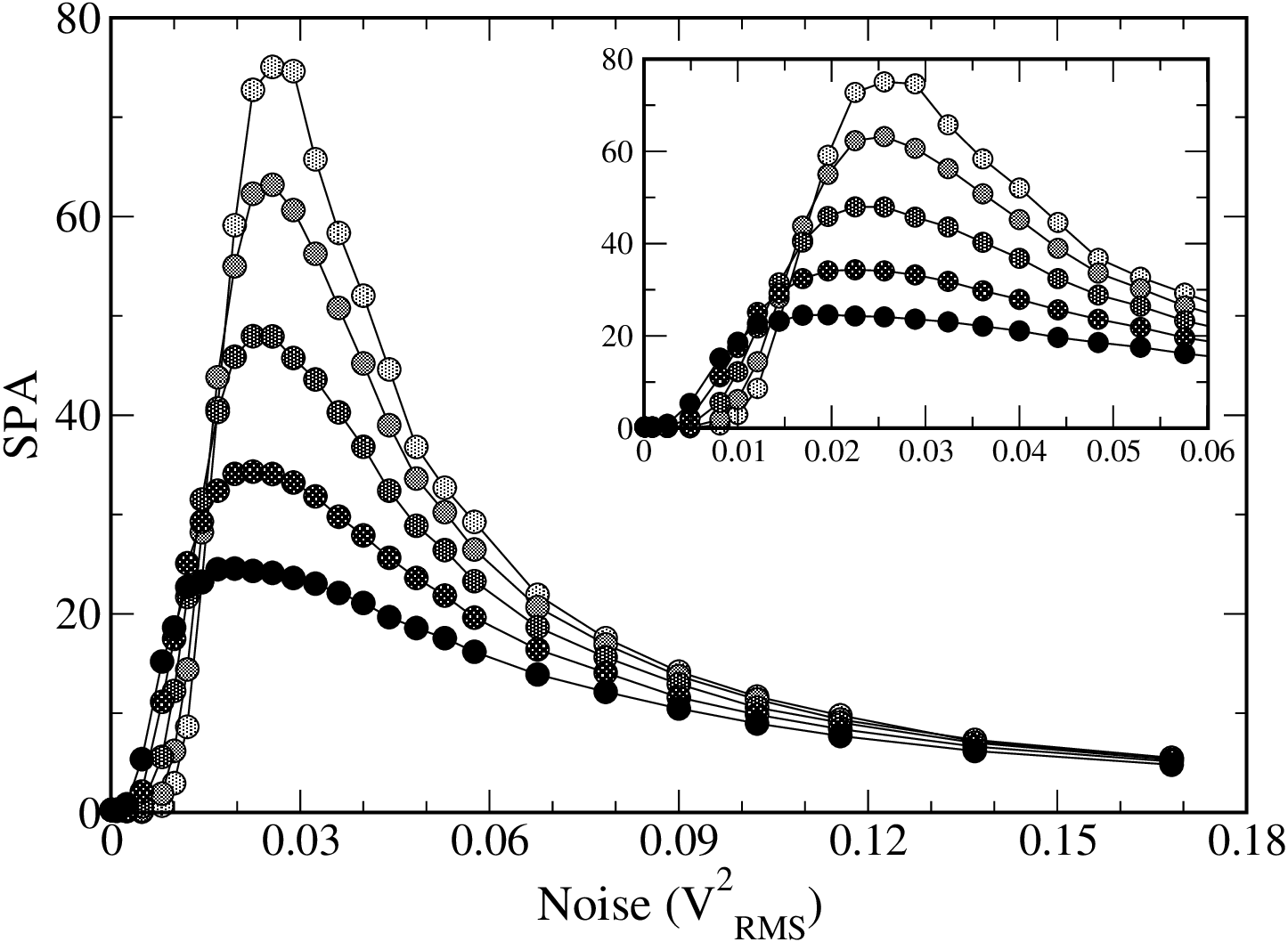}
\caption{The Spectral Power Amplification versus applied noise 
amplitude for different amplitudes of SIN modulation, 
$A_{SIN}=50, 100, 150, 200, 250$~mV (top to bottom). The inset 
is a magnification of the resonance region.}
\label{spa}
\end{figure}

To illustrate the phenomenon of SR, we first report in Fig.~\ref{spa} the
response of the system for different values of the amplitude of the SIN
modulation, as evaluated by means of the spectral power amplification  $SPA =
4\, |{\bf X}_{out}(\Omega)|^2/A_{SIN}^2$ \cite{jung1991}, where ${\bf
X}_{out}$ is the Fourier spectrum of the output signal evaluated at the 
driving frequency $\Omega=2\pi/T$. A well defined peak of SPA for an optimal
value of the applied noise is observed, yielding the typical signature of SR
\cite{rmp1998}. Increasing the input amplitude, a decrease in the response as
well as a shift towards lower noise values of the peak occurs, as already
reported from numerical simulations and observations in nonlinear circuits
\cite{nonlin}. 

The peak in the SPA indicates the improvement of the response of the system at
the modulation frequency, thus suggesting that a PS regime is achieved. To
investigate such a possibility we define the input--output
phase difference $\phi(t) \;=\; \Phi_{\rm out}(t) - \Phi_{\rm in}(t)$ where
$\Phi_{\rm in}(t)=\Omega t$. The instantaneous phase of a real signal $x(t)$
can be defined in a general way employing the concept of the analytic signal
\cite{bookpiko}. This is accomplished by considering the complex signal
$z(t)=x(t)+iy(t)$, where $y ={\bf H}x$ is the usual Hilbert transform of
$x(t)$. The phase is than given by the argument of $z$. For bistable systems an
alternative and easier definition can be given based on the knowledge of the
sequence of switching times $t_n$ only \cite{bookpiko}; $\Phi_{\rm
out}$ can be defined either as a piecewise--constant function increasing by
steps of $\pi$ at each jump time $t_n$ of the output or by linear interpolation.
In the analysis of experimental data we checked that different definitions
yield the same results, up to statistical accuracy.

\begin{figure}
\includegraphics[clip,width=4.2cm]{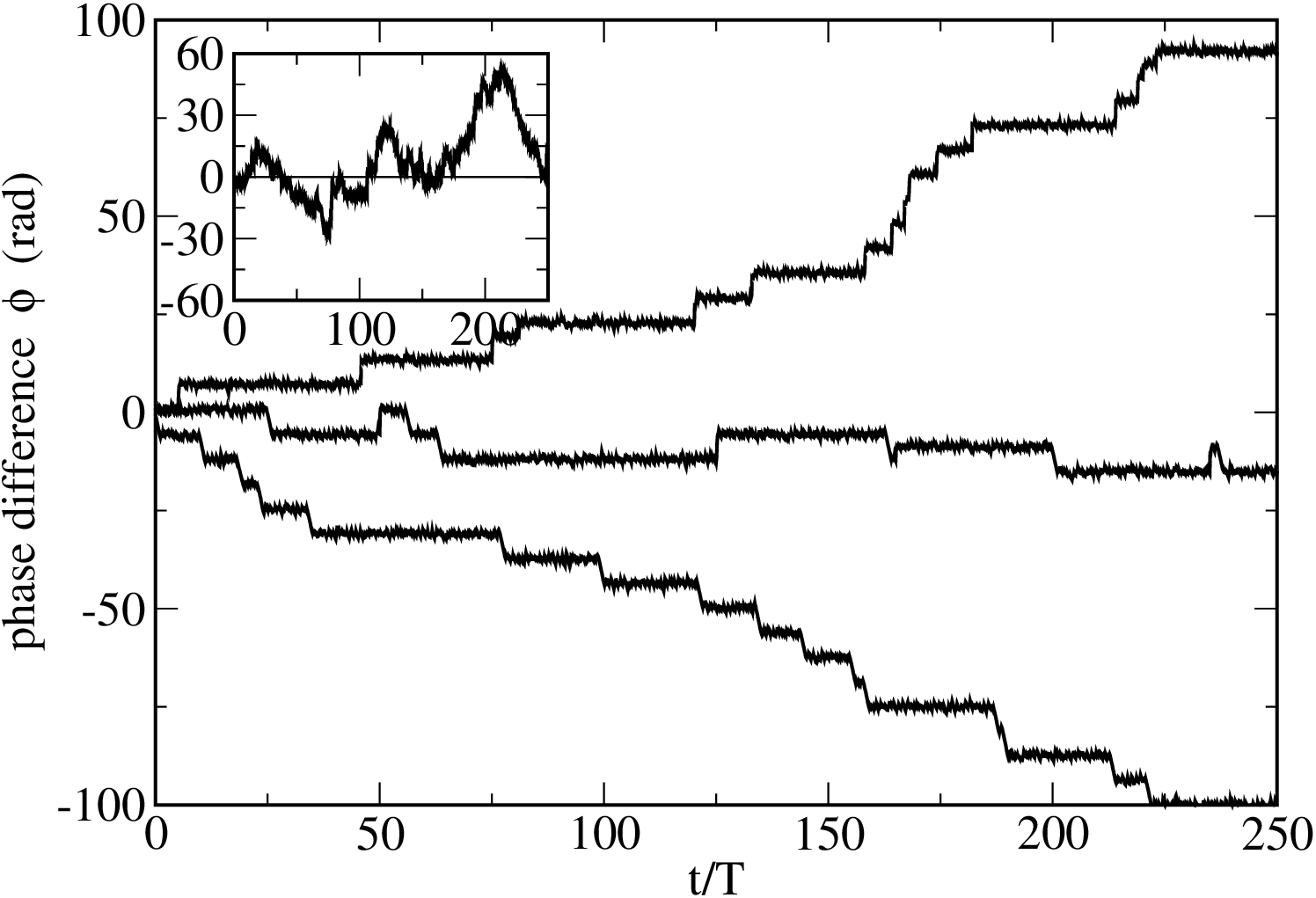}
\includegraphics[clip,width=4.2cm]{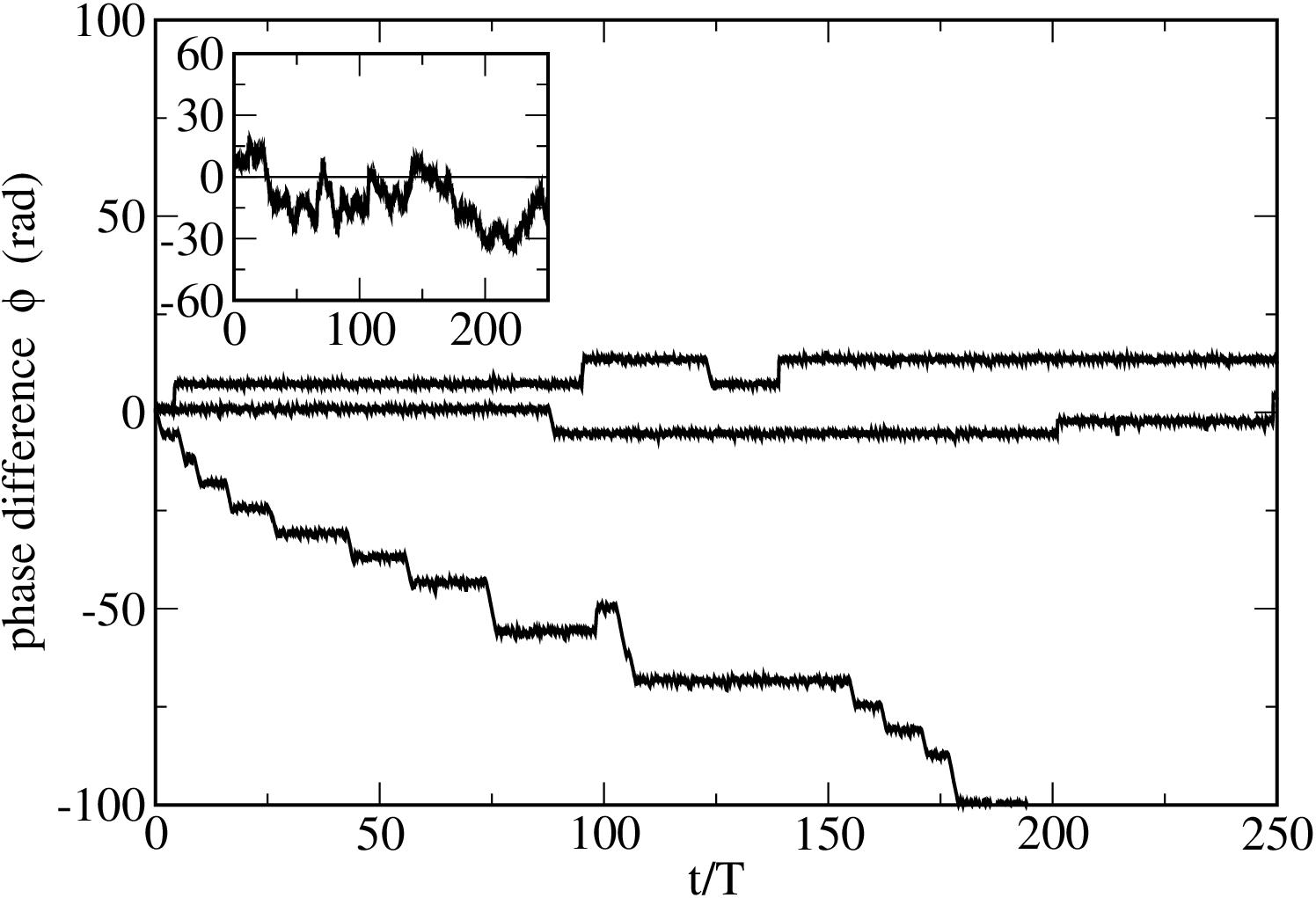}
\caption{The phase difference $\phi$  for SIN (left) and 
SQR (right) inputs of amplitude $A_{SIN}=250$~mV . Noise levels are 
0.0121, 0.0169, 0.0225 $V^2_{RMS}$ (bottom to top).
The insets are the time series for low-amplitude ($A_{SIN}=100$~mV) and 
noise level close to resonance (0.0196 $V^2_{RMS}$).} 
\label{temp}
\end{figure}

In Fig.~\ref{temp} we report the temporal evolution of $\phi$, as obtained from
the Hilbert transform method, for both the SIN and SQR waveforms and
different values of the added noise. In both cases, a clear signature of PS is
found at a well defined (resonant) noise power corresponding to the peak in the
SPA. As shown in the Figure, for a noise below (resp. above) the resonant value
the phase difference drifts towards lower (resp. higher) values. The
phase difference at resonance mantains an almost constant value, rarely
interrupted by jumps of integer multiples of $2\pi$ (phase slips).  For
comparison, in the insets of Figs.~\ref{temp} it is shown the temporal behavior
of $\phi$ at resonance for a smaller value of the input signal amplitudes; the
phase slips are here more frequent, yielding a faster diffusive motion. 

\begin{figure}
\includegraphics[clip,width=7.2cm]{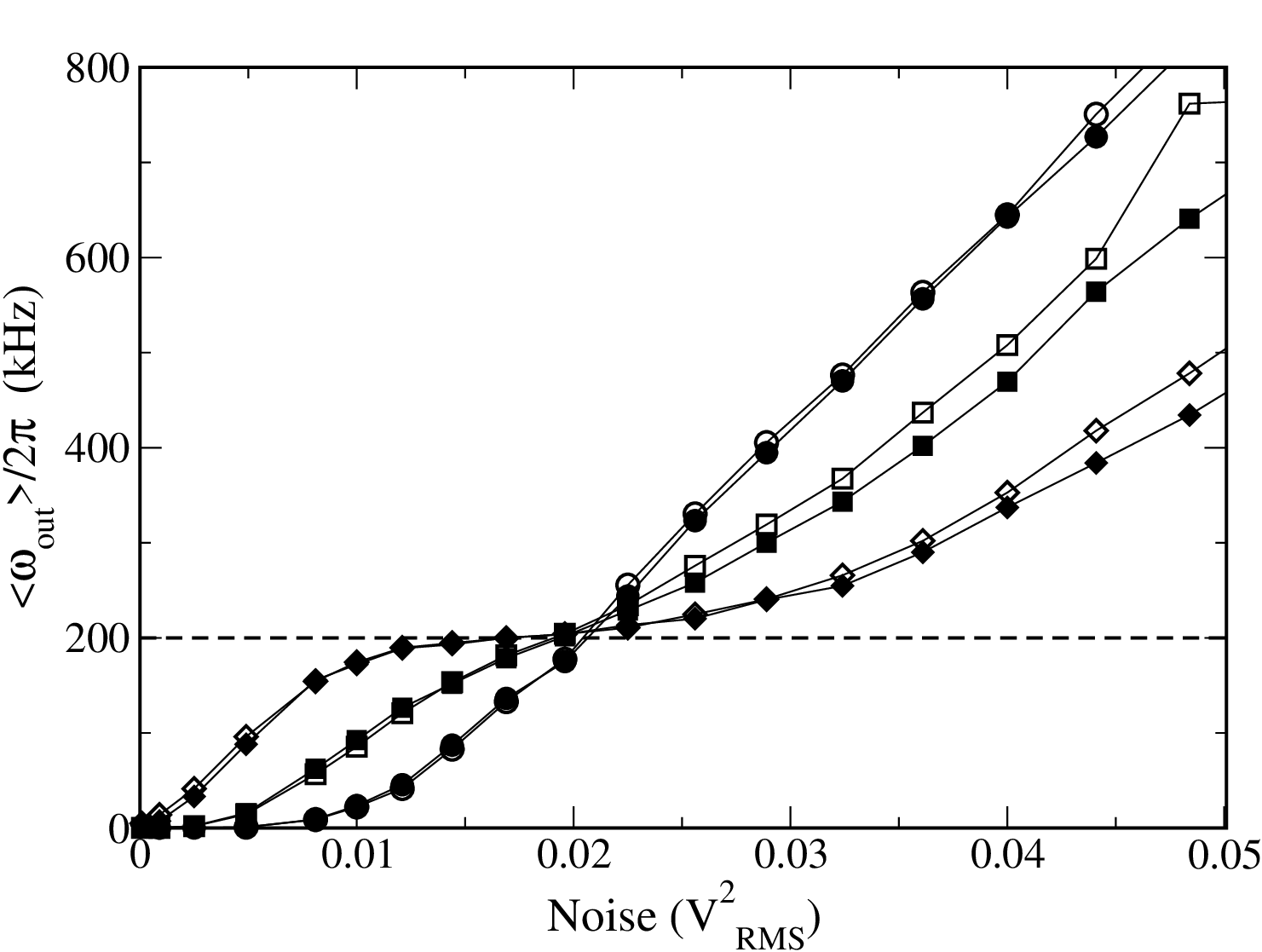}
\includegraphics[clip,width=7.2cm]{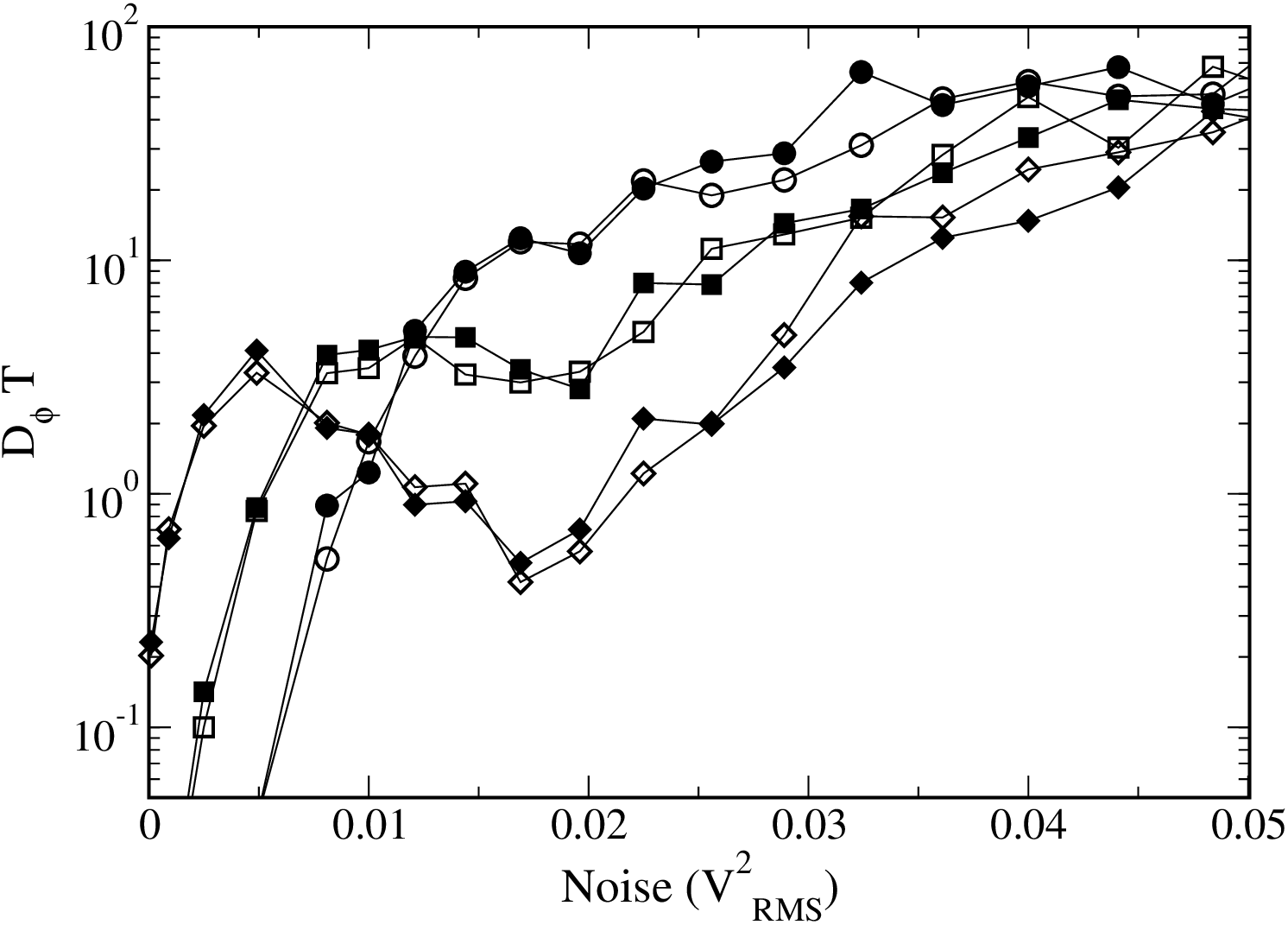}
\caption{Average output frequency (a) and diffusion coefficient (b)
for sinusoidal and square-wave inputs (full and open symbols respectively).
Input amplitudes are $A_{SIN}=50, 150, 250 \, mV$ (circles, squares, diamonds, 
respectively). 
} 
\label{moments}
\end{figure}

To quantify the observed behaviors, it is useful to introduce the average
frequency $\langle\omega_{\rm out}\rangle = \langle\dot{\Phi}_{\rm out}\rangle$ and
the effective diffusion coefficient $\deff \;=\; {1\over 2} {d \over dt}
\left[\langle\phi^2\rangle-\langle\phi\rangle^2\right]$ \cite{neim1998}. The
measured values of $\langle\omega_{\rm out}\rangle$ and $\deff$ are reported in
Fig.~\ref{moments} as a function of the noise intensity for different values of the
input signal amplitude $A$. Upon increasing $A$ the region of locking, i.e.  the
range of noise values for which  $\langle\omega_{\rm out}\rangle \simeq \Omega$,
widens while $\deff$ develops a pronouced dip at the resonant noise value. It is
worth noticing that, while the output frequency equals the input frequency at a
well defined value $D = D^*$,  increasing $A$ the
abscissa of the minimum of $\deff$ shifts towards lower noise amplitudes.
Moreover, such abscissa corresponds, within the  experimental accuracy, to the
abscissa of the maximum of the SPA.  Such result is in agreement with the analytic
predictions of Ref.\cite{freu2000}.

\begin{figure}
\includegraphics[clip,width=7.2cm]{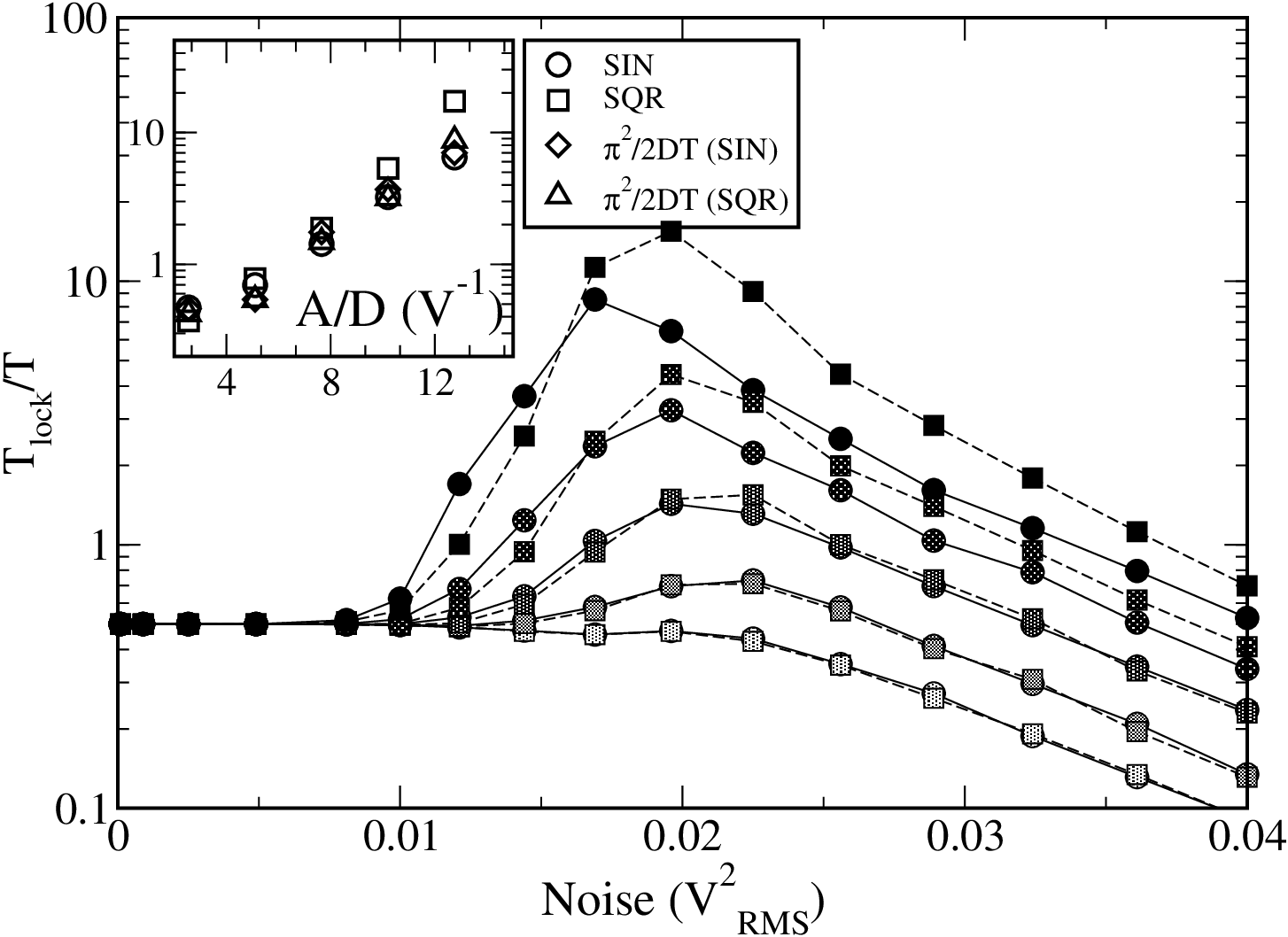}
\caption{Average duration of locking episodes 
for SIN and SQR inputs (circles and squares respectively), 
$A_{SIN}=50, 100, 150, 200, 250 \, mV$ 
(bottom to top). The inset shows $\tlock/T$ and $\pi^2/2\deff T$  
for $D=D^*=0.0196 V^2_{RMS}$) versus $A/D$ (see text).} 
\label{tlock}
\end{figure}

A related way to illustrate the enhacement of phase coherence is to  plot the
average duration of locking episodes $\tlock$ (i.e. the average time  between
consecutive phase slips) as a function of the noise \cite{freu2001}. As seen in
Fig.~\ref{tlock}, a marked maximum signals strong input-output correlations,
namely a decreasing rate of phase slips (see again Fig.~\ref{temp}). 
The minimum value of the diffusion constant $\deff$ and the maximal value of
$\tlock$ are indicators of the quality of PS. It is therefore important to
estimate their dependence on the experimental parameters. On the other 
hand, the two are not independent: as argued in Ref.~\cite{freu2001},  
the very definition of $\deff$ yields
\be
\langle \dot \phi \rangle ^2 \tlock^2 + 2 \deff \tlock \;=\; \pi^2 \quad.
\ee
For $D \simeq D^*$, we expect $\langle \dot \phi \rangle \simeq 0$, and therefore 
$\pi^2/2 \deff  \simeq  \tlock $. A shown in the inset of
Fig.~\ref{tlock}, this estimate is consistent with the experimental data for
both the input waveforms. 

A further relevant information is that $\tlock$ increases exponentially with
$A/D$, confirming that PS rapidly becomes effective upon increasing the driving
amplitude.  A simple argument to understand this scaling is the following.
Consider the case of a two--level system with a slowly modulated barrier
$\Delta V(t) = \Delta V \pm B$, where the sign changes every semiperiod
$T/2$.    The two Kramers' rates $a_{1,2} = r_K\exp(\mp B/D)$ give the
probability per unit time to jump from one level to the other. During the
locking period, the system stays synchronized with the modulation, likely
jumping (almost) every semiperiod. But the probability per unit time for the
system to jump less then (or more than) one time per semiperiod (resulting in a
phase slip) is approximatively $a_1$ for a sufficently large $B/D$. This
accounts for the observed scaling if $B$ is proportional to $A$. The same
conclusion can be drawn from Eq.~(\ref{eq:stochadler}). Close to resonance
($\Delta \simeq 0$)   $\deff \propto \exp(-const.\, \Delta_s/D)$
\cite{stra1967} and the exponential increase then follows from the fact that
$\Delta_s\propto A$ for small amplitudes.  

\begin{figure}[t]
\includegraphics[clip,width=7.cm]{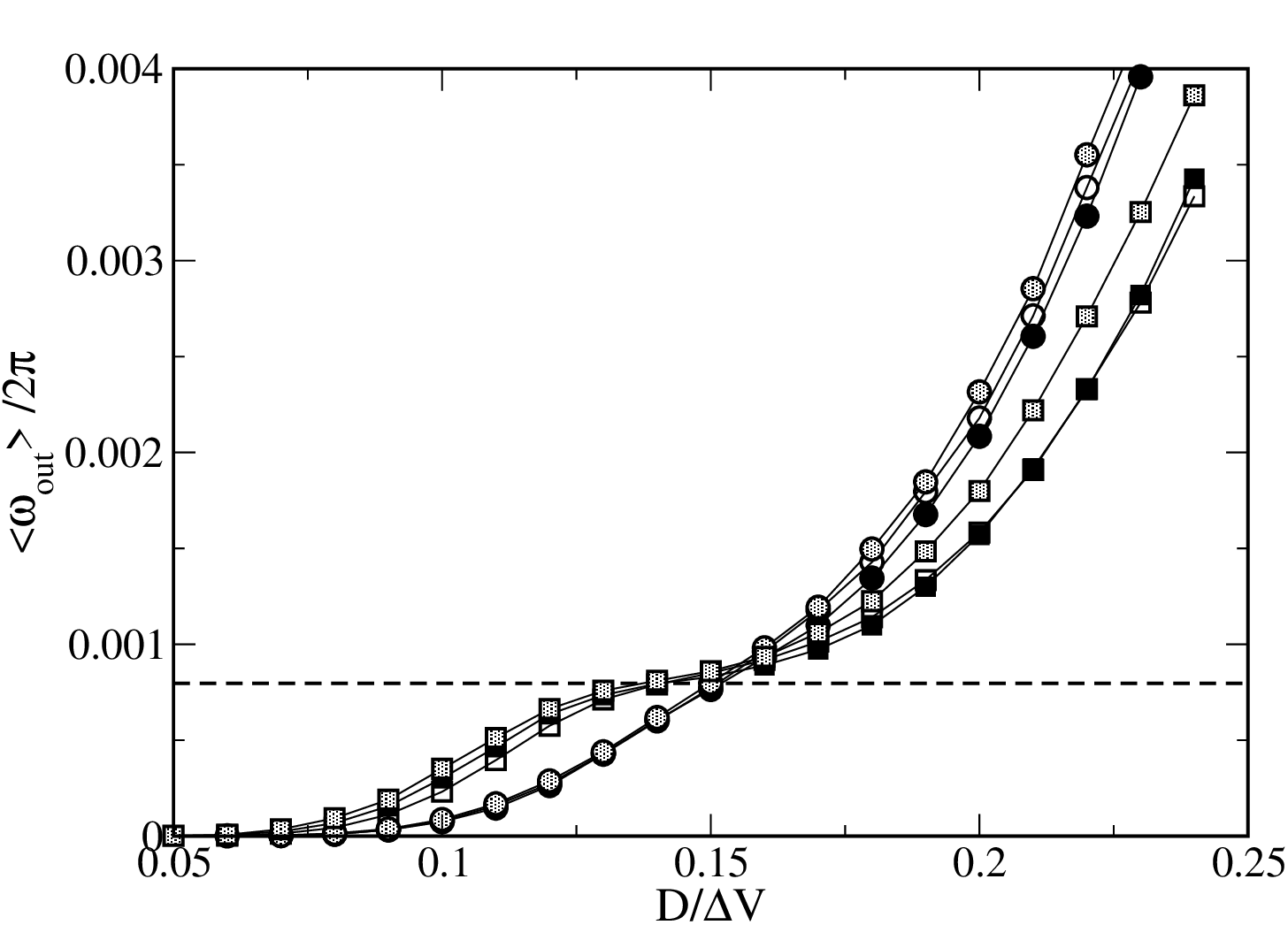}
\includegraphics[clip,width=7.cm]{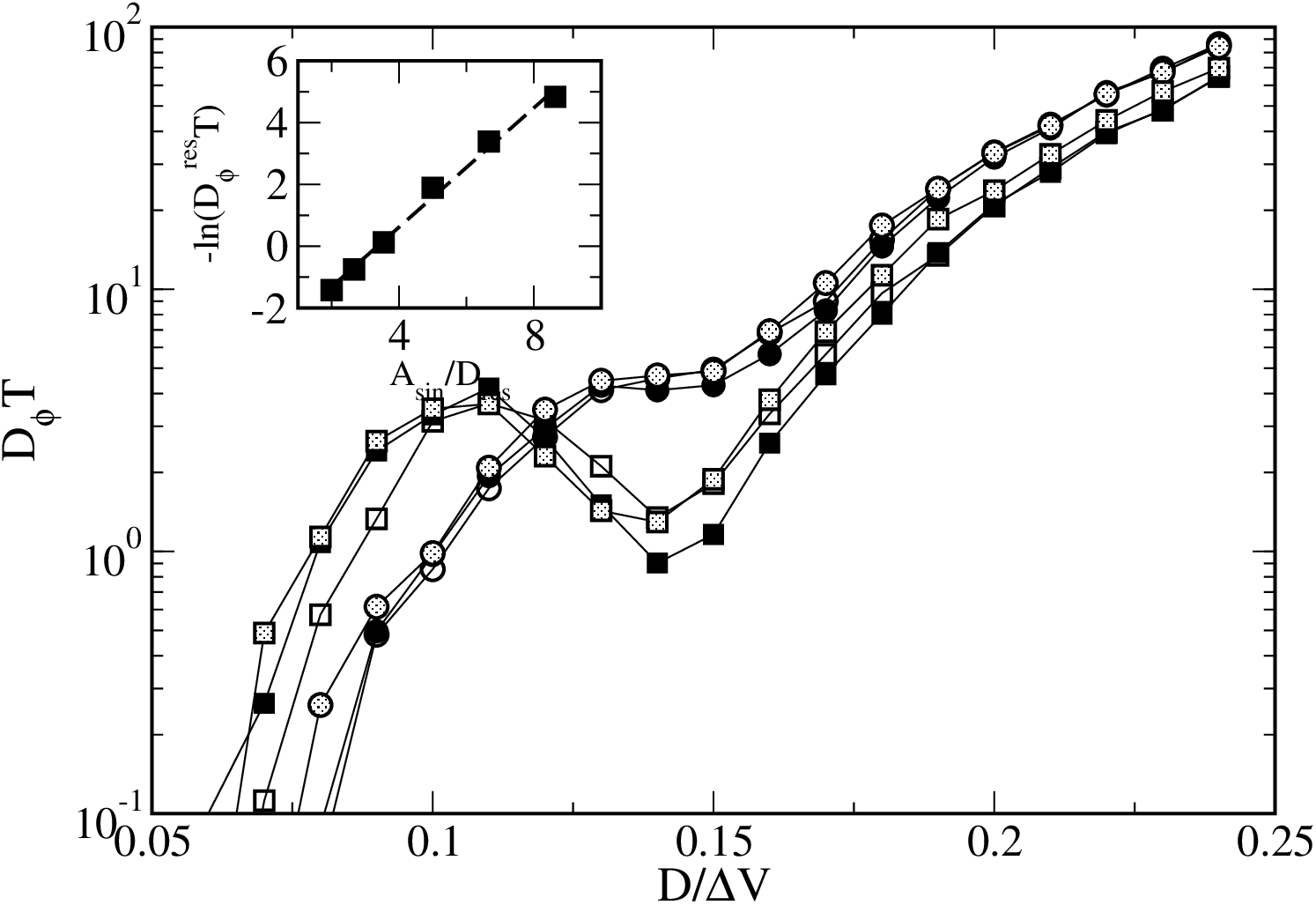}
\caption{Simulations of Eq.~(\ref{model}):
average output frequency and diffusion coefficient
for SIN, SQR and SAW inputs (full, open and grey 
symbols resp.) with $\Omega=0.005$, $A_{SIN}=0.3, 0.5$ (circles, squares 
resp.). 
} 
\label{lang}
\end{figure}

The experimental data are in qualitative agreement with  both the numerical
\cite{neim1998} and analytical \cite{freu2000} studies  of noise--induced PS.
This originates from the fact that, in the SR regime, the polarization dynamics
of our VCSEL can be succesfully modeled by the phenomenological Langevin
equation \cite{giac} \be \dot x = -V'(x) + Af(t) + \eta \quad, \label{model}
\ee  where $V$ is a double--well potential, $f$ is the modulation waveform (of
frequency $\Omega$) and $\eta$ is a white, Gaussian noise such that 
$\langle\eta(t)\eta(t')\rangle = 2 D \delta(t-t')$. Here the dimensionless
variable $x$ represents the polarized laser intensity. To emphasize the
correspondence  with the experiment we use here the same symbols $A,D,\Omega$
as above. However, one should be aware that this correspondence cannot be
established {\it a priori} since the model is only an effective description. A
direct quantitative comparison is feasible only by means of a careful
calibration of the physical parameters of model (\ref{model}) on the
experimental data. This task has been accomplished in slightly different 
experimental conditions \cite{giac}, and goes beyond the scope of the  present
work. Here, we limit ourselves to illustrate the results of simulations for the
model potential $V(x)=x^2(x^2-2)$ using a second--order stochastic Runge--Kutta
algorithm. The outcomes for moderate  values of $A$ ($A\lesssim
\Delta V$) shown in Fig.~\ref{lang} compare well with the experimental ones
(see again Figs.~\ref{moments}). Furthermore, the exponential scaling  
$\deff \propto \exp(-const.\, A/D)$ is found, in agreement with experimental
findings.

Let us now address the issue of how PS is affected by different choices of the
driving. Figs.~\ref{moments} indeed show that both $\langle\omega_{\rm
out}\rangle$ and $\deff$ are approximatively independent on the modulation
waveform, provided that  $A_{SQR}=A_{SIN}/\sqrt{2}$. The only relevant
parameter is,  at least in this range of amplitudes, the modulation frequency.
To check the generality  of this observation we performed simulation of
Eq.~(\ref{model}) with three different waveforms $f$, namely, a sinusoidal, a
square-wave and a sawtooth (SAW) with the same  RMS value, 
$A_{SQR}=A_{SIN}/\sqrt{2}=A_{SAW}/\sqrt{3}$.  As shown in Figs.~\ref{lang}, the
curves are almost overlapped indicating that for moderate amplitudes the
quality of PS is largely independent of the  details of the driving. To
understand this behaviour, we compared the Kramers' rates for the different
waveforms averaged on the modulation semiperiod. Upon changing $A/D$ in the
range of interest, we indeed found that they  are almost independent on $f$.
Therefore, in the adiabatic limit a comparison with the theory 
\cite{freu2000} can be  worked out and is currently in progress.

In conclusion, we reported a detailed experimental evidence of noise--induced
PS in a VCSEL. Phase entrainment is achieved by different input periodic
signals, with the same value for the measured statistical indicators as long as
the RMS of the signal is the same.  The numerical analysis of a Langevin model
reproduce the features observed experimentally, included the scaling of the
locking times at resonance with the amplitude of the modulation.

We thank S. Boccaletti for useful comments and for reading the manuscript.
This work is partially granted by the MIUR Project FIRB nr. RBNE01CW3M\_001.

\end{document}